\documentstyle[12pt,amsmath,amsfonts, amssymb, graphicx]{article}
\font\sd=cmcsc10.360pk

\numberwithin{equation}{section}

\newcommand{\lR}{\mathrm{I\hspace{-0.7mm}R}}
\newcommand{\lC}{\mathrm{l\hspace{-2.1mm}C}}

\begin{document}

\hoffset = -1truecm \voffset = -2truecm

\centerline{\Large\bf Flat BPS Domain Walls on $2d$ K\"ahler-Ricci
Soliton }

\medskip

\begin{center}
Bobby Eka Gunara and Freddy Permana Zen $^a$

\medskip

$^a$ Indonesia Center for Theoretical and Mathematical Physics (ICTMP)\\

and \\

Institut Teknologi Bandung, Indonesia\\
\end{center}

\bigskip

\textbf{Abstract.} In this paper we address several aspects of
flat Bogomolnyi-Prasad-Sommerfeld (BPS) domain walls together with
their Lorentz invariant vacua of $4d \; N=1$ supergravity coupled
to a chiral multiplet. The scalar field spans a one-parameter
family of $2d$ K\"ahler manifolds satisfying a K\"ahler-Ricci flow
equation. We find that BPS equations and the scalar potential
deform with respect to the real parameter related to the
K\"ahler-Ricci soliton. In addition, the analysis using gradient
and renormalization group flows are carried out to ensure the
existence of Lorentz invariant vacua related to Anti de
Sitter/Conformal Field Theory (AdS/CFT) correspondence.


\section{Introduction}
Deformation of $N=1$ BPS supergravity domain walls on a
K\"ahler-Ricci soliton, which generates a one-parameter family of
K\"ahler manifolds, has been firstly studied in \cite{GZ, GZ1} by
us. In the future we plan to apply this work to study the Anti de
Sitter/Conformal Field Theory (AdS/CFT) correspondence in four
dimensions and its evolution on a K\"ahler-Ricci soliton.\\

\indent The aim of this paper is to consider a two dimensional
model of $4d \; N=1$ supergravity BPS domain walls which
generalizes a simple $\;{\mathrm{\lC P}}^1$ model given in
\cite{GZ} as an example and the static case \cite{GZA}. The $N=1$
theory is coupled to a chiral multiplet where its complex scalar
spans a one-parameter family of $2d$ K\"ahler manifolds generated
by the $2d$ K\"ahler-Ricci flow equation \cite{cao}\footnote{See
also \cite{topping, caozhu} for an excellent review of this
subject.}
\begin{equation}
\frac{\partial g_{z\bar{z}}}{\partial \tau} = -2
R_{z\bar{z}}(\tau) = -2 \;\partial_z
\bar{\partial}_{\bar{z}}{\mathrm{ln}}g_{z\bar{z}}(\tau)\;,
\label{KRF2d}
\end{equation}
whose solutions can be regarded as an area deformation of a
K\"ahler surface for finite time. This can be easily seen, for
example, when the initial geometry is a K\"ahler-Einstein surface
\begin{equation}
R_{z\bar{z}}(0) = \Lambda_2 \, g_{z\bar{z}}(0) \;, \label{KEM}
\end{equation}
where $g_{z\bar{z}}(0) >0$ and $\Lambda_2 > 0$. Then, in this case
the solution of (\ref{KRF2d}) is given by
\begin{equation}
g_{z\bar{z}}(\tau) = (1-2\Lambda_2 \tau) \, g_{z\bar{z}}(0) \;,
\label{KRFsolKE}
\end{equation}
and the area of the soliton is
\begin{equation}
\vert 1-2\Lambda_2 \tau \vert \sqrt{{\mathrm{det}}g(0)} \, dz
d\bar{z} \;.
\end{equation}
Clearly, the soliton in (\ref{KRFsolKE}) admits a singularity at
$\tau = 1/ 2\Lambda_2$. For $0 \le \tau < 1/2\Lambda_2$, the
geometry is diffeomorphic to the initial geometry, while after
hitting the singularity the new geometry has
$\hat{g}_{z\bar{z}}(0) < 0$ and $\hat{\Lambda}_2 < 0$. This
example shows that the $2d$ K\"ahler-Ricci soliton indeed defines
a family of  K\"ahler surfaces.\\

\indent In this paper we consider a case where the walls preserve
half of the supersymmetry of the parental theory described by a
set of BPS equations.  We use dynamical system analysis to study
the BPS equations and the renormalization group (RG) flow equation
to determine properties of the critical points describing Lorentz
invariant vacua and their deformation.\\

\indent Some results can be shortly mentioned as follows. The
K\"ahler-Ricci soliton does play a role in determining the
stability of the walls near critical points of the BPS equations
describing Lorentz invariant vacua. As we will see, this occurs
because the geometric (K\"ahler-Ricci) flow controls the signature
of a K\"ahler metric. The analysis fails when the critical points
become degenerate which might be a bifurcation point. Furthermore,
the RG flow analysis shows that those vacua do not always exist
particularly in the infrared (low energy scale) regions. It is
important to note that the geometric flow may diverge at finite
$\tau$, where $\tau$ is a real parameter related to the
K\"ahler-Ricci flow. In this situation, some quantities blow up.\\

 \indent The organization of this paper is as
follows. In section 2 we give a quick introduction of $4d$ N=1
supergravity on $2d$ K\"ahler-Ricci soliton. Then we discuss some
properties of flat BPS domain wall solutions of the theory in
section 3. Section 4 provides a discussion about some aspects of
Lorentz invariant vacua, gradient  and renormalization group (RG)
flows. We then apply the analysis to two simple models in section
5. We summarize the results in section 6.

\section{$N=1$ Chiral Supergravity on $2d$ K\"ahler-Ricci Soliton}
\label{CSKRS}

In this section we review the four dimensional $N=1$ supergravity
coupled to a chiral multiplet in
 which the non-linear $\sigma$-model can be viewed as a one-parameter
family of $2d$ K\"ahler manifolds generated by the K\"ahler-Ricci
flow equation (\ref{KRF2d}) and deformed with respect to the real
parameter $\tau$ \cite{GZ, GZ1}. The structure of this section
follows rather closely
reference \cite{GZ1}.\\

The building blocks of the $N=1$ theory are a gravitational
multiplet and a chiral multiplet. The field components of the
gravitational multiplet are a vierbein $e^a_{\nu}$ and a vector
spinor $\psi_{\nu}$ where $a=0,...,3$ and $\nu=0,...,3$ are the
flat and the curved indices, respectively. The chiral multiplet
consists of a complex
scalar $z$ and a spin-$\frac{1}{2}$ fermion $\chi$.\\

\indent Then, a general $N=1$ chiral supergravity Lagrangian
together with its supersymmetry transformation can be constructed.
A pedagogical review of this subject can be found, for example, in
\cite{susy}. Here, we assemble the terms which are useful for our
analysis. The bosonic part of the $N=1$ supergravity Lagrangian
has the form
\begin{equation}
{\mathcal{L}}^{N=1} = -\frac{M^2_P}{2}R + g_{z\bar{z}}(z, \bar{z};
\tau)\,
\partial_{\nu} z \,
\partial^{\nu}\bar{z} - V(z,\bar{z}; \tau)\;, \label{L}
\end{equation}
where $M_P$ is the Planck mass \footnote{Setting  $M_P \to
+\infty$, one gets the global $N=1$ chiral supersymmetric
theory.}, $R$ is the Ricci scalar of the four dimensional
spacetime; and the pair $(z,\bar{z})$ spans a Hodge-K\"ahler
manifold with metric $g_{z\bar{z}}(z, \bar{z}; \tau) \equiv
\partial_{z}
\partial_{\bar{z}}K(z,\bar{z}; \tau)$; and $K(z,\bar{z}; \tau)$
is a real function, called the K\"ahler potential. The $N=1$
scalar potential $V(z,\bar{z}; \tau)$ has the form
\begin{equation}
 V(z,\bar{z}; \tau) = e^{K(\tau)/M^2_P}\left(g^{z\bar{z}}(\tau)\nabla_z W\,
 \bar{\nabla}_{\bar{z}} \bar{W}
 - \frac{3}{M^2_P} W \bar{W} \right)\;,
\label{V}
\end{equation}
where $W$ is a holomorphic superpotential and $\nabla_z W\equiv
(dW/dz)+ (K_z(\tau)/M^2_P) W$. The supersymmetric invariance of
the Lagrangian (\ref{L}) is guaranteed by the following
transformations of the
 fields up to three-fermion terms \footnote{The symbol
$D_{\nu}$ here is different with the one in reference \cite{GZA}.
$D_{\nu}$ here is defined as $D_{\nu} \equiv \partial_{\nu}
-\frac{1}{4}\gamma_{ab}\,\omega^{ab}_{\nu}$.}
\begin{eqnarray}
\delta\psi_{1\nu} &=& M_P \left(D_{\nu}\epsilon_1 +
\frac{\mathrm{i}}{2}e^{K(\tau)/2M^2_P}\,W \gamma_{\nu} \epsilon^1
+ \frac{\mathrm{i}}{2M_P} Q_{\nu}(\tau)\epsilon_1 \right)\;, \nonumber\\
\delta\chi^z &=& {\mathrm{i}}\partial_{\nu} z \, \gamma^{\nu}
\epsilon^1 + N^z(\tau) \epsilon_1 \quad, \label{susytr}\\
\delta e^a_{\nu} &=&  - \frac{{\mathrm{i}}}{M_P} \, (
\bar{\psi}_{1\nu} \, \gamma^a \epsilon^1
+ \bar{\psi}^1_\nu \, \gamma^a \epsilon_1 )\;,\nonumber\\
\delta z &=& \bar{\chi}^z \epsilon_1 \;,\nonumber
\end{eqnarray}
where $N^z(\tau) \equiv
e^{K(\tau)/2M^2_P}\,g^{z\bar{z}}(\tau)\bar{\nabla}_{\bar{z}}\bar{W}$,
 $g^{z\bar{z}}(\tau) = (g_{z\bar{z}}(\tau))^{-1}$, and the $U(1)$ connection
 $Q_{\nu}(\tau) \equiv  - \left(  K_z(\tau) \,\partial_{\nu}z -
K_{\bar{z}}(\tau)\, \partial_{\nu}\bar{z}\right)$. Here, we have
also introduced $\epsilon_1 \equiv \epsilon_1(x,\tau)$. One can
then show that using (\ref{KRF2d}), the evolution equations of
both the scalar potential (\ref{V}) and the shifting quantity
$N^z$ can be written as \cite{GZ}
\begin{eqnarray}
\frac{\partial N^z(\tau)}{\partial \tau} &=& 2 R^z_{\;z}(\tau)
N^z(\tau)+ \frac{K_{\tau}(\tau)}{2M_P^2} N^z(\tau)+
g^{z\bar{z}}(\tau) \frac{K_{\bar{z}\tau}(\tau)}{M_P^2}
e^{K(\tau)/2M^2_P}\, \bar{W}\;, \nonumber\\
 \frac{\partial V(\tau)}{\partial
\tau} &=& \frac{\partial N^z(\tau)}{\partial \tau} N_z(\tau)+
\frac{\partial N_z(\tau)}{\partial \tau} N^z(\tau) -
\frac{3K_{\tau}(\tau)}{M_P^2} e^{K(\tau)/M^2_P}\, \vert W \vert^2
\;, \nonumber
\end{eqnarray}
where $R^z_{\;z} \equiv g^{z\bar{z}} R_{z\bar{z}}$.\\

\indent It is important to remark, as mentioned in the previous
section, that the metric $g_{z\bar{z}}(\tau)$ could possibly turn
into a negative definite metric. Such a case would yield a
non-unitary theory with ghosts. At the classical level, it may be
possible to formally write down theories of this type. However, at
the quantum level this would lead to negative norm states in the
standard Fock space which are discarded as unphysical \footnote{We
thank M. Zagermann for pointing out these aspects.}.

\section{Flat BPS Domain Walls}

Now we give a discussion about the ground states which partially
break Lorentz invariance, \textit{i.e.}, domain walls. The first
step is to take the ansatz metric as
\begin{equation}
ds^2 = a^2(u,
\tau)\,\eta_{\underline{\lambda}\,\underline{\nu}}\,dx^{\underline{\lambda}}
dx^{\underline{\nu}}-du^2 \quad,\label{DWans}
\end{equation}
where $\underline{\lambda}, \,\underline{\nu}=0,1,2$, $a(u, \tau)$
is the warped factor, and
$\eta_{\underline{\lambda}\,\underline{\nu}}$ is the flat
Minkowskian metric. Writing the supersymmetry transformation
(\ref{susytr}) on the background (\ref{DWans}) and setting
$\psi_{1\mu}=\chi^z=0$, then a set of equations that partially
preserves supersymmetry can be obtained in the following
\begin{eqnarray}
\frac{a'}{a}  &=& \pm \, {\mathcal{W}}(\tau) \;, \nonumber\\
 z' &=& \mp 2 g^{z\bar{z}}(\tau)
\bar{\partial}_{\bar{z}}{\mathcal{W}}(\tau) \; , \label{gfe}\\
\bar{z}' &=& \mp 2 g^{z\bar{z}}(\tau)
\partial_z{\mathcal{W}}(\tau) \;,\nonumber
\end{eqnarray}
where
\begin{eqnarray}
a' &\equiv& \partial a/\partial u \;,  \nonumber\\
{\mathcal{W}}(\tau) &\equiv& e^{K(\tau)/M^2_P}\, \vert W \vert
\;,\label{curlW}
\end{eqnarray}
and we have assumed $\epsilon_1(u, \tau)$ and $z(u, \tau)$. The
last two equations in (\ref{gfe}) are called BPS equations.
Moreover, in the conformal field theory (CFT) picture one can then
define a beta function for all scalar fields as
\begin{equation}
 \beta \equiv a \frac{\partial z}{\partial a} = -
 2g^{z \bar{z}}(\tau) \frac{\bar{\partial}_{\bar{z}} {\mathcal{W}}}{\mathcal{W}} \quad,
\label{beta}
\end{equation}
 after using  (\ref{gfe}), together with its complex conjugate describing another
 supersymmetric RG flows. In this picture the scalars can be viewed  as
  coupling constants and the warp factor $a$ is playing the role
 of an energy scale \cite{DWc, CDKV}. It is worth mentioning that equation (\ref{gfe}) can also
be derived in more general four dimensional gravity theory without
introducing supersymmetry (\ref{susytr}). In this case one has to
solve the Einstein's field equations together with the equation of
motions of the complex scalar fields. These equations can be
constructed by varying the Lagrangian (\ref{L}) with respect to
the spacetime metric $g_{\mu\nu}$ and the scalar field $z$, and
also, applying the necessary condition that the variation of
$\tau$ vanishes. As a result, we find that the real function
${\mathcal{W}}(\tau)$ has a general form and
\begin{equation}
{\mathcal{W}}(\tau) \ne e^{K(\tau)/M^2_P}\, \vert W \vert \;.
\label{nosusy}
\end{equation}
The corresponding scalar potential is given by
\begin{equation}
 V(z,\bar{z}; \tau) = 4\, g^{z\bar{z}}(\tau)\, \partial_z {\mathcal{W}}(\tau)\,
 \bar{\partial}_{\bar{z}} {\mathcal{W}}(\tau)
 - \frac{3}{M^2_P}\, {\mathcal{W}}^2(\tau) \;.
\label{V1}
\end{equation}
 Then, the CFT
picture can also be defined by the beta function (\ref{beta}) in a
more general way, namely equation (\ref{nosusy}) holds.\\
 \indent Finally, it is straightforward to show that the
critical points of the BPS equations in (\ref{gfe}) are related to
the following condition
\begin{equation}
\partial_z{\mathcal{W}} = \bar{\partial}_{\bar{z}}{\mathcal{W}}
=0 \quad, \label{critpo}
\end{equation}
which implies that
\begin{equation}
\partial_z V = \bar{\partial}_{\bar{z}} V =0 \quad. \label{critpo0}
\end{equation}
This means that the critical points of ${\mathcal{W}}(z,\bar{z})$
are somehow related to the critical points of the $N=1$ scalar
potential $V(z,\bar{z})$. Moreover, in the view of (\ref{beta}),
these points are in ultraviolet (UV) region if $a \to \infty$ and
in infrared (IR) region if $a \to 0$. Thus, the RG flow
interpolates between UV and IR critical points.

\section{Supersymmetric Vacua and AdS/CFT Correspondence}

Let us begin our discussion by mentioning that from (\ref{critpo})
a critical point of the real function ${\mathcal{W}}(\tau)$, say
$p_0$, is in general $p_0 \equiv (z_0(\tau), \bar{z}_0(\tau))$ due
to the geometric flow (\ref{KRF2d}). Such a point exists in the
asymptotic regions, namely around $u \to \pm \infty$. The form of
the scalar potential (\ref{V}) at $p_0$ is
\begin{equation}
 V(p_0; \tau) = - \frac{3}{M^2_P}\, {\mathcal{W}}^2(p_0; \tau)
 \equiv - \frac{3}{M^2_P}\, {\mathcal{W}}^2_0\;.
\label{V2}
\end{equation}
For the case of the flat domain walls discussed in \cite{GZA}
equation (\ref{V2}) can be viewed as the cosmological constant
  of the spacetime at the vacuum. So the spacetime is AdS with
  negative cosmological constant for ${\mathcal{W}}_0 \ne 0$.
  Here, we exclude the case of Minkowskian vacua, where ${\mathcal{W}}_0 =
  0$.\\

\indent The next step is to consider some aspects of the critical
points of ${\mathcal{W}}(\tau)$ and the vacua of the theory
defined by the critical points of the scalar potential (\ref{V}).
First, we write down the eigenvalues of ${\mathcal{W}}(\tau)$ and
the scalar potential in (\ref{V}) as
\begin{eqnarray}
 \lambda^{\mathcal{W}}_{1,2}(\tau) &=& \frac{g_{z\bar{z}}(p_0; \tau)}
 {M_P^2}{\mathcal{W}}_0
 \pm 2 \lvert \partial^2_z {\mathcal{W}}_0\rvert  \:, \nonumber\\
 \lambda^{V}_{1,2}(\tau) &=& -4 \left( \frac{g_{z\bar{z}}(p_0; \tau)}{M_P^4}
 {\mathcal{W}}_0^2 - 2 g^{z\bar{z}}(p_0; \tau) \lvert \partial^2_z
 {\mathcal{W}}_0 \rvert^2 \right)\nonumber\\
 && \pm \, 4 \frac{{\mathcal{W}}_0}{M_P^2} \lvert \partial^2_z
 {\mathcal{W}}_0\rvert \:,
 \label{eigenval}
 \end{eqnarray}
respectively, where
\begin{equation}
\partial^2_z {\mathcal{W}}_0 \equiv \frac{e^{K (p_0; \tau)/M^2_P}\,
\bar{W}(\bar{z}_0)}{2{\mathcal{W}}(p_0; \tau)} \left(
\frac{d^2W}{dz^2}(z_0) + \frac{K_{zz}(p_0; \tau)}{M_P^2}W(z_0)+
\frac{K_z(p_0; \tau)}{M_P^2}\frac{dW}{dz}(z_0)\right)\:.
\label{duaW}
\end{equation}
 The analysis of the above eigenvalues has been carried out in
\cite{GZ1, GZA}. Here, we just summarize it in the following
table.

\medskip
\begin{flushleft}
Table 1 Deformation of Vacua due to a K\"ahler-Ricci Soliton
\end{flushleft}
\begin{tabular}{|l|c|l|l|}
\hline
K\"ahler Metric & Condition & Critical Points of
${\mathcal{W}}(\tau)$ & Type
of Vacua  \\
 \hline \hline
 &  &  &  local minimum  \\
 & $\lvert \partial^2_z
{\mathcal{W}}_0\rvert > \frac{g_{z\bar{z}}(p_0; \tau)}{2M_P^2}
   {\mathcal{W}}_0 $ & saddle  & saddle \\
$g_{z\bar{z}}(\tau) > 0$ & & & degenerate \\ \cline{2-4}
 & $\lvert \partial^2_z{\mathcal{W}}_0\rvert < \frac{g_{z\bar{z}}(p_0; \tau)}{2M_P^2}
   {\mathcal{W}}_0 $ & local minimum  &  local maximum \\ \cline{2-4}
& $\lvert \partial^2_z{\mathcal{W}}_0\rvert =
\frac{g_{z\bar{z}}(p_0; \tau)}{2M_P^2}
   {\mathcal{W}}_0 $ & degenerate  &  intrinsic degenerate \\
 \hline \hline
&  &  &  local maximum  \\
 & $\lvert \partial^2_z
{\mathcal{W}}_0\rvert > \frac{g_{z\bar{z}}(p_0; \tau)}{2M_P^2}
   {\mathcal{W}}_0 $ & saddle  & saddle \\
$g_{z\bar{z}}(\tau) < 0$ & & & degenerate \\ \cline{2-4}
 & $\lvert \partial^2_z{\mathcal{W}}_0\rvert < \frac{g_{z\bar{z}}(p_0; \tau)}{2M_P^2}
   {\mathcal{W}}_0 $ &  local maximum &  local minimum \\ \cline{2-4}
& $\lvert \partial^2_z{\mathcal{W}}_0\rvert =
\frac{g_{z\bar{z}}(p_0; \tau)}{2M_P^2}
   {\mathcal{W}}_0 $ & degenerate  &  intrinsic degenerate \\
\hline
\end{tabular}

\bigskip
As we saw in the table, it is possible to have a parity
transformation of the vacua caused by the K\"ahler-Ricci soliton.
This can be easily seen in the local minimum vacua in which they
are changing to a local maximum vacua as the metric deformed to be
a negative definite metric.\\

 \indent Furthermore, we have to check the stability of the walls near $p_0$
 by expanding the BPS equations in
 (\ref{gfe}) near $p_0$ to first order. It turns out that the
first order expansion matrix has the eigenvalues
\begin{equation}
 \Lambda_{1,2}=  \mp \frac{{\mathcal{W}}_0}{M_P^2}- 2 g^{z\bar{z}}(p_0; \tau)
 \lvert \partial^2_z {\mathcal{W}}_0\rvert
 \label{eigenf} \:,
\end{equation}
which describe the gradient flows. If $g^{z\bar{z}}(p_0; \tau)$ is
positive definite for finite $\tau$, then the stable flow is a
stable node flowing along local minimum vacua and the stable curve
of the saddle vacua, whereas the unstable flow is a saddle flowing
along the local maximum vacua. If $g^{z\bar{z}}(p_0; \tau)$
becomes negative definite for finite $\tau$, then we have an
unstable node flowing along the local maximum vacua and the
unstable direction of the saddle vacua, while the stable saddle
flow is flowing along the local minimum vacua. These show that in
the latter case the walls are mostly unstable, and they are stable
if the gradient flow is the stable saddle flow. In addition, Our
linear analysis fails if one eigenvalue in (\ref{eigenf}) is zero
which is possibly a fold bifurcation point. This happens at
degenerate critical points of the real function which
correspondingly, are related to the degenerate critical points of
the scalar potential
(\ref{V}).\\

 \indent Now we turn to consider the RG flows.
 As we have mentioned in the preceding section,
this function can also be used to determine the nature of the
critical point $p_0$, namely, it can be interpreted as the UV or
IR limit in the CFT picture. To begin, we expand the beta function
(\ref{beta}) to first order around $p_0$ and then we have a matrix
whose eigenvalues are
\begin{equation}
\lambda^{\mathcal{U}}_{1,2} = \frac{1}{M_P^2} \pm \frac{2
g^{z\bar{z}}(p_0; \tau) }{{\mathcal{W}}_0}\,\lvert
\partial^2_z {\mathcal{W}}_0\rvert\:\,. \label{eigenU}
\end{equation}
Let us choose a model where the UV region is $u \to + \infty$ as
$a \to + \infty$, while the IR region is $u \to - \infty$ as $a
\to 0$. In the UV region at least one of the eigenvalues (12)
should be positive. Moreover, since it is possible to have zero
and a negative eigenvalue, i.e. for minus sign, as the parameters
in ${\mathcal{W}}(\tau)$ or the flow parameter $\tau$ vary, the RG
flow fails to depart the UV region in this direction. So, the flow
is stable along the positive eigenvalue and we have all possible
vacua
in the UV region.\\

\indent On the other hand, in the IR region the RG flow approaches
a critical point in the direction of negative eigenvalue. However,
since this eigenvalue could be zero and positive as the parameters
vary, then the IR critical points would vanish. In the scalar
potential picture there are no IR local maximum vacua for the
positive definite K\"ahler metric, {\textit{i.e.}}
$g_{z\bar{z}}(\tau) > 0$, while no IR local minimum vacua for the
negative definite K\"ahler metric, {\textit{i.e.}}
$g_{z\bar{z}}(\tau) < 0$. In addition, for both cases intrinsic
degenerate vacua do not exist in this IR region.\\

\indent It is important to note that the above analysis fail if
there exists a singularity of the K\"ahler-Ricci soliton at a
particular value of $\tau$.

\section{Two $\;{\mathrm{l\hspace{-2.9mm}C}}{\mathrm{P}}^1(p_0)$ Models}

In this section we organize the discussions into two parts. The
first part is dealing with a linear  superpotential, while the
second part is a model with harmonic superpotential. Moreover,
this section is devoted for considering a complex projective
manifold with the center $p_0 \equiv (z_0, \bar{z}_0) \ne 0$ is
denoted by $\;{\mathrm{\lC P}}^1 (p_0)$ and the K\"ahler potential
is given by
\begin{equation}
K(0) = {\mathrm{ln}}(1+\vert z - z_0\vert^2) \:, \label{KCP0}
\end{equation}
where $p_0$ is a vacuum and for $p_0 = 0$ we just write
$\;{\mathrm{\lC P}}^1$. Then the $\tau$-dependent K\"ahler
potential is given by
\begin{equation}
K(\tau) = \sigma(\tau) \, {\mathrm{ln}}(1+\vert z - z_0\vert^2)
\:, \label{KCP1}
\end{equation}
where $\sigma(\tau) \equiv (1-4\tau)$. For $\tau < 1/4$ we have a
manifold which \textbf{is} diffeomorphic to $\;{\mathrm{\lC P}}^1 (p_0)$,
while after hitting the singularity at $\tau = 1/4$ we have $\;
\widehat{{\mathrm{\lC P}}^1}(p_0)$ which is called a parity
manifold of $\;{\mathrm{\lC P}}^1 (p_0)$.

\subsection{A Model with Linear Superpotential}

In this subsection we consider a model with linear superpotential
\begin{equation}
W(z) = a_0 + a_1 \, (z - z_0) \:, \label{W}
\end{equation}
where $a_0, a_1 \in \lR$. Equation (\ref{critpo}) in this case
becomes
\begin{equation}
 a_1 + \frac{\sigma(\tau)(\bar{z} - \bar{z}_0)}{M_P^2(1+\vert z - z_0\vert^2)}
 (a_0 + a_1 \, (z - z_0)) = 0\:. \label{critpo1}
\end{equation}
Taking $z = z_0$, it follows that $a_1 =0$ and the ground state is
an unstable isolated AdS spacetime with the cosmological constant
\begin{equation}
V(p_0; \tau) = - \frac{3|a_0|^2}{M_P^2} \:,
\end{equation}
for $a_0 \ne 0$ and describing unstable walls for $\tau < 1/4$.
The ground state becomes stable for $\tau
> 1/4$.\\
\indent If there exists another vacuum $p'_0 \ne p_0$, by defining
$z'_0 \equiv x'_0 + {\mathrm{i}} y'_0$ with $x'_0, y'_0 \in \lR$,
then the condition (\ref{critpo1}) gives
\begin{eqnarray}
x'_0(\tau) &=& \frac{1}{2(\sigma(\tau)+M_P^2)}\!\left\lbrack
{-}\frac{\sigma(\tau) a_0 }{a_1}
\pm\!\left(\!\left(\frac{\sigma(\tau) a_0 }{a_1}\right)^2\,{-}\,4
M_P^2
(\sigma(\tau)+M_P^2) \right)^{\!\!1/2} \right\rbrack + x_0 \;, \nonumber\\
 y'_0(\tau) &=& y_0 \;, \label{vacuum}
\end{eqnarray}
with $a_1 \ne 0$. Moreover, since $x'_0, y'_0 \in \lR$ the flow
parameter $\tau$ must be defined in the following interval
\cite{GZ}
\begin{eqnarray}
\tau & \le & \frac{1}{4} - \frac{M_P^2}{2} \left(\frac{a_1
}{a_0}\right)^2
 \left\lbrack
1 + \left( 1 + \left( \frac{a_0}{a_1}\right)^2 \right)^{1/2}
\right\rbrack\!, \nonumber\\
\tau & \ge & \frac{1}{4} + \frac{M_P^2}{2} \left(\frac{a_1
}{a_0}\right)^2 \left\lbrack -1 + \left(1 + \left(
\frac{a_0}{a_1}\right)^2 \right)^{1/2} \right\rbrack\!,
\label{welldef}
\end{eqnarray}
before and after the singularity at $\tau = 1/4$, respectively.
For the shake of simplicity we take $a_0 \gg a_1 > 0$. Afterward,
we have the same discussion as in \cite{GZ}. Here we summarized
the results in the following table.
\medskip
\begin{flushleft}
Table 2 Deformation of Vacua in a $\;{\mathrm{\lC P}}^1 (p_0)$
Model with Linear Superpotential
\end{flushleft}
\begin{tabular}{|l|c|l|l|l|}
\hline
K\"ahler Metric & Interval & Type
of Vacua & Region & Stability \\
\hline \hline
 &  $\tau <  - \frac{a_0 M_P^2}{8 a_1}$ &  local maximum & UV & unstable \\ \cline{2-5}
 &  $- \frac{a_0
M_P^2}{8 a_1}  < \tau <   - \frac{a_0 M_P^2}{8 a_1 \sqrt{2}}$ &
saddle & UV, IR & stable \\ \cline{2-5}
 $g_{z\bar{z}}(p_0; \tau) > 0$ & $-
\frac{a_0 M_P^2}{8 a_1 \sqrt{2}} < \tau \lesssim \frac{1}{4}
  - \frac{a_1 M_P^2}{2a_0}$ & local minimum & UV, IR & stable \\ \cline{2-5}
 $0 \le \tau < 1/4$ & $\tau \approx  - \frac{a_0 M_P^2}{8 a_1 \sqrt{2}}$  & degenerate
 & UV, IR & stable \\ \cline{2-5}
& $\tau \approx - \frac{a_0 M_P^2}{8 a_1}$ &  intrinsic degenerate & UV & undetermined\\
 \hline \hline
& $\frac{1}{4} + \frac{a_1 M_P^2}{2a_0} \lesssim \tau < \frac{a_0
M_P^2}{8 a_1 \sqrt{2}}$ &  local maximum &  UV, IR & unstable\\
\cline{2-5}
 & $\frac{a_0 M_P^2}{8 a_1 \sqrt{2}}   < \tau <
 \frac{a_0 M_P^2}{8 a_1}$ & saddle & UV, IR & unstable \\ \cline{2-5}
$g_{z\bar{z}}(p_0; \tau) < 0$ & $\tau >  \frac{a_0 M_P^2}{8 a_1}$
& local minimum & UV & stable \\ \cline{2-5}
$ \tau > 1/4$ & $\tau \approx  \frac{a_0 M_P^2}{8 a_1 \sqrt{2}}$ &
degenerate & UV, IR & unstable \\ \cline{2-5}
 & $\tau \approx  \frac{a_0 M_P^2}{8 a_1}$ & intrinsic degenerate & UV & undetermined\\
\hline
\end{tabular}

\bigskip
Our comments are in order. As we have seen in the above table, all
possible cases may exist in the UV region. However in the IR
region, this is not the case. Before the singularity at $\tau =
1/4$ the local maximum vacua do not exist in the IR region, while
after the singularity, the local minimum vacua are forbidden. In
both cases, there are no intrinsic degenerate vacua in this IR
region.


\subsection{A Model with Harmonic Superpotential}

In this subsection we discuss properties of a model with the
harmonic superpotential
\begin{equation}
W(z) = A_0 \, {\mathrm{sin}} (kz)  \:, \label{Wharm}
\end{equation}
where $A_0$ and $k$ are real \cite{GZA}. Similar to the preceding
subsection, taking $z = z_0$, the condition (\ref{critpo}) results
in
\begin{equation}
  y_0 = 0 \:,\: x_0 = (n+ \frac{1}{2})\frac{\pi}{k} \:, \quad n = 0, \pm 1, \pm 2,
  ....,
\end{equation}
where we have also defined $z_0 \equiv x_0 + {\mathrm{i}} y_0$.
For the case at hand, the ground state is an AdS spacetime for
$A_0 \ne 0$ with
\begin{equation}
V(p_0; \tau) = - \frac{3|A_0|^2}{M_P^2} \:.
\end{equation}
One can further obtain
\begin{eqnarray}
\lambda^{\mathcal{W}}_{1,2} &=& \left( \frac{\sigma(\tau)}{M_P^2} \pm k^2 \right)|A_0| \:,\nonumber\\
\lambda^{V}_{1,2} &=& 2|A_0|^2 \left( \sigma(\tau)^{-1}\, k^4  -
\frac{2\sigma(\tau)}{M_P^4} \pm \frac{k^2}{M_P^2}\right) \:,
\end{eqnarray}
from (\ref{eigenval}) and in this model the possible bifurcation
points occur at $k = \pm \vert \sigma(\tau) \vert^{1/2} / M_P$ and
$\tau \ne 1/4$.\\
\indent Finally we want to mention that if there exists another
vacuum $z'_0 \ne z_0$, we will get an implicit equation which is
complicated to have its exact solution. This aspect will be
discussed elsewhere.

\section{Conclusions}
We have studied general properties of the BPS domain walls on a
$2d$ K\"ahler-Ricci soliton. The critical points of the scalar
potential describing the supersymmetric vacua do deform with
respect to the real parameter $\tau$ related to the geometric
soliton.\\

\indent Firstly, for $g_{z\bar{z}}(\tau) > 0$ with finite $\tau$,
the real function ${\mathcal{W}}(\tau)$ have three types of
critical points, namely the local minima, the saddles, and the
degenerate critical points. The local minima correspond to the
local maximum vacua, whereas the saddles are mapped into a local
minimum or saddle vacua. The degenerate critical points imply
intrinsic degenerate vacua. On the other side, if the metric
$g_{z\bar{z}}(\tau)$ becomes negative definite for finite $\tau$,
then the local minima of ${\mathcal{W}}(\tau)$ turn into a local
maxima. At the level of vacua such situation could possibly occur.
This shows the existence of the parity transformation for the
critical points or the vacua of the theory.\\

\indent The analysis using gradient flows shows that if the metric
$g_{z\bar{z}}(\tau) > 0$ for finite $\tau$, then we have stable
nodes flowing along the local minimum vacua and the stable curve
of the saddle vacua. The other flows are unstable saddles flowing
along local maximum vacua. Next, if $g_{z\bar{z}}(\tau) < 0$ for
finite $\tau$, then there are unstable nodes flowing along the
local maximum and the unstable direction of the saddle vacua,
whereas the stable saddle flows are flowing along the local
minimum vacua. These facts show that as mentioned above, there
exists a possibility of having a parity transformation of gradient
flows caused by the soliton. In addition, the analysis is failed
when the critical points of ${\mathcal{W}}(\tau)$ become
degenerate which is the bifurcation point of the
gradient flows.\\

\indent On the other side, the RG flows determine whether such
gradient flows exist in the UV and IR regions. In the UV region
all possible vacua are allowed for all value of
$g_{z\bar{z}}(\tau)$. However, such situation does not valid in
the IR regions. In the case where $g_{z\bar{z}}(\tau) > 0$, we do
not have IR local maximum vacua, whereas for the case
$g_{z\bar{z}}(\tau) < 0$ there are no IR
local minimum vacua.\\

\indent At the end we applied the general analysis to two simple
models on $\;{\mathrm{\lC P}}^1 (p_0)$. If we take $z = z_0$, then
the cosmological constants do not depend on $\tau$. However, the
second order analysis does depend on $\tau$ which shows that the
properties of the ground states may be changed because of the
K\"ahler-Ricci soliton. If there is another vacuum $z'_0 \ne z_0$,
then the case becomes complicated. Thus, for example it is easier
for the case of linear superpotential.

\section*{Acknowledgment}
We thank L. Andrianopoli, H. Alatas, Arianto, A. N. Atmaja, K.
Yamamoto, J. M. Maldacena, T. Mohaupt,  and M. Zagermann for
useful discussions. One of us (BEG) is grateful to the organizer
of the Abdus Salam ICTP Spring School on Superstring Theory and
Related Topics for providing a good atmosphere where parts of this
work was done. We finally thank M. Satriawan for careful reading
the manuscript and correcting the grammar. This work was initiated
by Riset KK ITB 2008 under contract No. 035/K01.7/PL/2008 and is
extensively supported by Hibah Kompetensi DIKTI 2008 No.
040/HIKOM/DP2M/2008 and Hibah Kompetensi DIKTI 2009 No.
223/SP2H/PP/DP2M/V/2009.

\newpage

\vspace{1cc}

{\small\noindent { \sd Bobby Eka Gunara}\\
Indonesia Center for Theoretical and
Mathematical Physics (ICTMP),
\\
and \\
Theoretical  Physics Laboratory, \\
Theoretical High Energy Physics and Instrumentation Division,\\
Faculty of Mathematics and Natural Sciences, \\
Institut Teknologi Bandung \\
Jl. Ganesha 10 Bandung 40132, Indonesia.,\\
\vspace{0.5cc} \noindent \hspace*{-0.25cm} E-mail:
bobby@fi.itb.ac.id}

{\small\noindent { \sd Freddy Permana Zen}\\
Indonesia Center for Theoretical and
Mathematical Physics (ICTMP),
\\
and \\
Theoretical  Physics Laboratory, \\
Theoretical High Energy Physics and Instrumentation Division,\\
Faculty of Mathematics and Natural Sciences, \\
Institut Teknologi Bandung \\
Jl. Ganesha 10 Bandung 40132, Indonesia.,\\
\vspace{0.5cc} \noindent \hspace*{-0.25cm} E-mail:
fpzen@fi.itb.ac.id}

\end{document}